\documentclass[aps,prl,twocolumn,preprintnumbers,reprint,floatfix,nofootinbib]{revtex4-1}
\pdfoutput=1
\usepackage{graphicx}
\usepackage{bm}
\usepackage{times}
\usepackage{slashed}
\usepackage{color}
\usepackage{slashed}
\usepackage{lipsum}
\usepackage{subfigure}
\usepackage[utf8]{inputenc}
\usepackage{amsfonts}
\usepackage{mathrsfs}
\usepackage{amsmath}
\usepackage{graphicx,epsfig}
\usepackage{url}
\usepackage{amssymb}
\usepackage{comment}
\newcommand{\be}{\begin{equation}}
\newcommand{\ee}{\end{equation}}
\newcommand{\bea}{\begin{eqnarray}}
\newcommand{\eea}{\end{eqnarray}}
\usepackage{hyperref}
\usepackage{amsmath}
\usepackage{graphicx}
\usepackage[colorinlistoftodos]{todonotes}
\newcommand{\mplanck}{M_{pl}}
\newcommand{\mdm}{M_{DM}}
\newcommand{\cw}{\mathrm{c}_W^2}
\newcommand{\sw}{\mathrm{s}_W^2}
\newcommand{\cwf}{\mathrm{c}_W^4}
\newcommand{\swf}{\mathrm{s}_W^4}

\begin{document}

\title{Gamma-ray lines may reveal the CP nature of the dark matter particle}

\author{Farinaldo S. Queiroz}
\email{farinaldo.queiroz@iip.ufrn.br}

\affiliation{International Institute of Physics, Universidade Federal do Rio Grande do Norte,
Campus Universitario, Lagoa Nova, Natal-RN 59078-970, Brazil}

\author{Carlos E. Yaguna}
\email{carlos.yaguna@uptc.edu.co}
\affiliation{
Escuela de Física, Universidad Pedagógica y Tecnológica de Colombia (UPTC),\\
Avenida Central del Norte, Tunja, Colombia}

\begin{abstract}
Determining the fundamental properties of the dark matter is one of the most important open problems in particle physics today. If the dark matter particle has spin zero, one of these properties is its CP nature. That is, whether it is CP-even (scalar), CP-odd (pseudoscalar),  or if its interactions  violate CP. In this paper, we show that the observation of $\gamma$-ray lines arising from the \emph{decay} of  a spin-zero dark matter particle could be used to discriminate among these possibilities. We consider a general setup where dark matter decay is induced by  effective operators  and demonstrate that, due to gauge invariance, there exists correlations among the branching ratios into gauge boson final states ($\gamma\gamma$, $\gamma Z$, $W^+W^-$, $ZZ$) that depend on the mass and the CP properties of the dark matter. Consequently, the future observation of  $\gamma$-ray lines  may in principle be used to establish the CP nature of the dark matter particle.
\end{abstract}


\maketitle

\section{Introduction}
Even though dark matter accounts for about $25\%$ of the energy density of the Universe \cite{Ade:2015xua},  we know almost nothing about it. A new elementary particle, not currently known and not included in the Standard Model,  is required to explain this exotic form of matter, but it is not at all clear what particle it is. In fact, one of the most important open problems in particle physics today is precisely that of identifying the dark matter particle or, equivalently, determining its fundamental properties. 

To that end, the first step would be the detection of the dark matter particle by non-gravitational means. Currently, dark matter is being searched for in direct and indirect dark matter detection experiments \cite{Undagoitia:2015gya,Aprile:2018dbl,Strigari:2013iaa,Ackermann:2015zua,Baring:2015sza}, as well as at the LHC \cite{Kahlhoefer:2017dnp,Aaboud:2017phn}. 
A  dark matter signal may be detected anytime, so it is crucial to be prepared to extract from it as much information as possible about the nature of the dark matter particle. Several studies have been carried out along these lines --see e.g. \cite{Arina:2013jya,Ferrer:2015bta,Anderson:2015xaa,Rogers:2016jrx,Roszkowski:2016bhs,Catena:2017xqq,Bertone:2017adx}. They generally aim at measuring the dark matter mass and the relevant cross section (or lifetime), which would certainly help identifying the dark matter particle.   But couldn't we also  unveil some of its fundamental properties? 
Very few works have  addressed this important issue.  In \cite{Queiroz:2016sxf,Kavanagh:2017hcl} it was shown that it may be possible to determine whether the dark matter particle is identical to its antiparticle while  in \cite{Catena:2017wzu,Catena:2018uae}  it was demonstrated that the spin of the dark matter particle could be revealed. These analyses were all based on the observation of dark matter signals in future direct detection experiments. 

In this work we address for the first time the possibility of determining the CP properties of the dark matter particle.  We show that it is in principle feasible to find out whether the dark matter particle is a scalar, a pseudoscalar,  or if its interactions  violate CP. Our proposal relies  on the observation of $\gamma$-ray lines arising from the \emph{decay} of  a spin-zero dark matter particle ($\varphi\to\gamma\gamma$ and $\varphi\to\gamma Z$), and on the correlation between their branching ratios, which  is enforced by gauge invariance. Specifically, we demonstrate that, within a generic framework in which dark matter decay is induced by effective operators, such correlation exists and depends on the mass and the CP nature of the dark matter. Thus, the future observation of  $\gamma$-ray lines may be used to establish whether the dark matter particle is CP-even (scalar), CP-odd (pseudoscalar), or  CP-violating, providing a decisive clue towards the identification of  the dark matter particle.

\section{Framework}

\begin{figure*}[tbh]
\includegraphics[width=0.9\columnwidth]{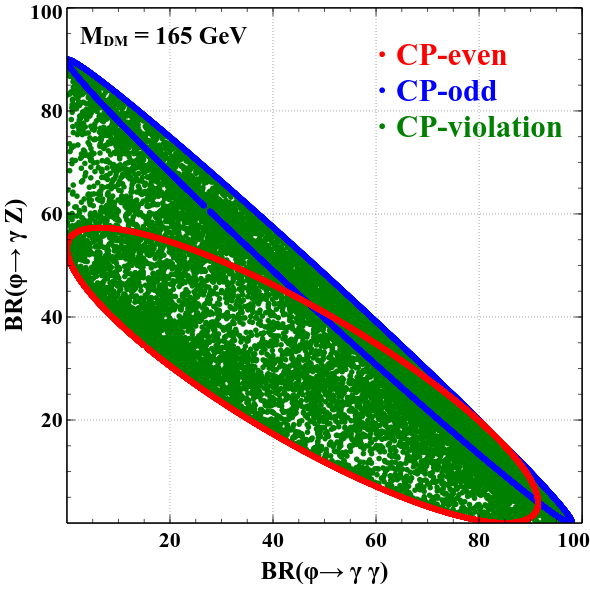}
\hspace{5mm}
\includegraphics[width=0.9\columnwidth]{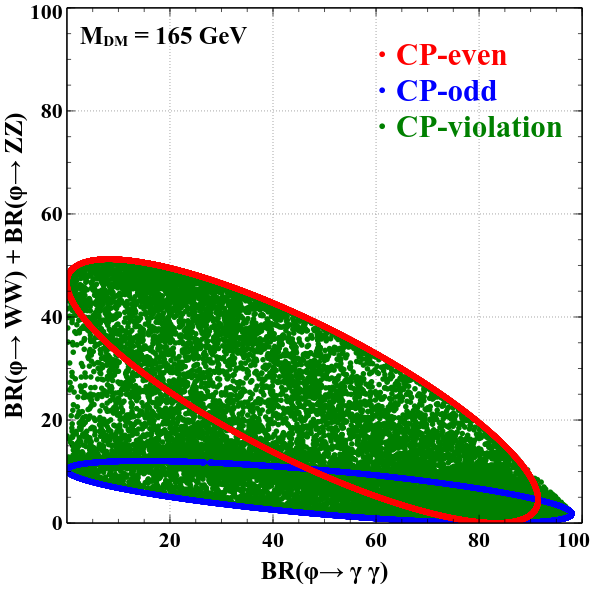}
\caption{Correlations between the branching ratios into $\gamma Z$, $WW$ and $\gamma\gamma$ for  $M_{DM}=165~$GeV.\label{fig:corr1}}
\end{figure*}
 Let us assume that future indirect detection experiments observe a  $\gamma$-ray signal whose morphology is consistent with  \emph{decaying} \cite{Bertone:2007aw,Cirelli:2012ut,Ibarra:2013cra,Slatyer:2016qyl} rather than annihilating dark matter. From a theoretical point of view,  such a signal can be explained within the framework of effective field theory. At the end, the dark matter lifetime must be much larger than the age of the Universe \cite{Mambrini:2015sia}, a fact that could be attributed to the scale suppression associated with higher dimensional effective operators. For definiteness, we will consider  a dark matter particle, $\varphi$, that has  spin zero and is a singlet under the SM gauge group and assume that the detected dark matter signal features some \emph{prominent} $\gamma$-ray lines. $\gamma$-ray lines have received a lot of attention in indirect detection analyses \cite{Yaguna:2009cy,Weniger:2012tx,Aisati:2015ova,Ackermann:2015lka} because, being essentially free of astrophysical backgrounds, they provide a smoking-gun signature of dark matter. The most general effective  Lagrangian that can induce the decay of $\varphi$ into $\gamma$-ray lines  is 
\begin{align}
\label{eq:Lagrangian}
\mathscr{L}&= \frac{a}{\mplanck}\varphi B_{\mu\nu}B^{\mu\nu}+\frac{b}{\mplanck}\varphi B_{\mu\nu}\tilde{B}^{\mu\nu}\nonumber\\
& +\frac{c}{\mplanck}\varphi W^I_{\mu\nu}W^{I\mu\nu}+\frac{d}{\mplanck}\varphi W^I_{\mu\nu}\widetilde{W}^{I\mu\nu}
\end{align}
where $\mplanck$ is the Planck mass (an arbitrary scale $\Lambda$ may  be used instead),  $a,b,c,d$ are  dimensionless coupling constants, and $B_{\mu\nu}$ and $W^I_{\mu\nu}$ are the strength-field tensors associated with $U(1)_Y$ and $SU(2)$, respectively, and  $\tilde{B}_{\mu\nu}$ and $\widetilde{W}^I_{\mu\nu}$ their corresponding dual tensors. Notice that all these operators have mass dimension 5.  There are no operators at dimension 6 inducing the decay --the only possibility is the Weinberg operator multiplied by $\varphi$, which does not produce  $\gamma$-ray lines-- whereas  at dimension 7, the operators that may produce lines are all suppressed by the factor $(v/\mplanck)^2$  with respect to those we consider.  The $\gamma$-ray lines could also be generated through loop-diagrams, e.g. via $\varphi$ coupling to SM fermions,  but in that case  they are  expected to be significantly  suppressed with respect to the continuum contribution, a fact that would hinder the possibility of actually observing them.

 Given this interaction Lagrangian, the CP nature of the dark matter  is determined by the couplings $a,b,c,d$ as follows: if $b=d=0$, only the CP-even terms survive and we say that $\varphi$ is a proper scalar (or CP-even); if $a=c=0$, only the CP-odd terms survive and $\varphi$ is  a pseudoscalar (or CP-odd); in any other situation we would simultaneously have CP-even and CP-odd terms, which would imply CP violation. 

Hence, one way in which the CP nature of the dark matter could be determined is by extracting these couplings directly from the observed $\gamma$-ray spectrum. For instance, if a fit to  the spectrum were to require $a,d\neq 0$, CP violation would have been established. We advocate, however, a different approach based on the correlations among the branching ratios into different final states. The above Lagrangian induces the decay of $\varphi$ into the following two-body  final states\footnote{The decay into the three-body final states $W^+W^-\gamma$ and $W^+W^-Z$, which are induced by the $SU(2)$ terms, are suppressed and will be neglected in the following.}: $\gamma\gamma$, $\gamma Z$, $W^+W^-$, and $ZZ$. The novel idea of this work is that, due to gauge invariance, there exists correlations among the branching ratios into these four final states  that depend on the CP nature of the dark matter. Thus, the  detection of  signals consistent with the decay of a scalar dark matter particle into  $\gamma$-ray lines could be used to determine whether the dark matter particle is CP-even, CP-odd or CP-violating.

\section{Results}

Let us now illustrate how such correlations differ according to the CP nature of the dark matter particle (see the appendix for the analytic expressions required to produce the figures). Figure \ref{fig:corr1} shows, for $\mdm=165~$GeV, scatter plots of $BR(\varphi\to \gamma\gamma)$ versus $BR(\varphi\to \gamma Z)$ in the left panel and of $BR(\varphi\to \gamma\gamma)$ versus $BR(\varphi\to W^+W^-)+BR(\varphi\to ZZ)$ in the right panel\footnote{Since $\mdm=165~$GeV in this case, this sum actually correspond to just $BR(\varphi\to W^+W^-)$.}.  Three sets of points are shown in each panel, corresponding to the possible CP properties of the dark matter. The points denoted by CP-even (in red) were obtained by randomly varying $a$ and $c$ while setting $b=d=0$.  The points denoted by CP-odd (in blue) were obtained instead by randomly varying $b$ and $d$ while setting $a=c=0$. Finally, the points denoted by CP violation (in green) correspond to randomly varying all four couplings --$a,b,c,d$. Notice that the CP-even and the CP-odd points lie along different ellipses on both planes, whereas the points featuring CP violation  are spread over a much larger area.  These figures already demonstrate the main point of this paper: the observation of such decays may provide direct information on the CP nature of the dark matter. 

\begin{figure}[tbh!]
\includegraphics[width=0.9\columnwidth]{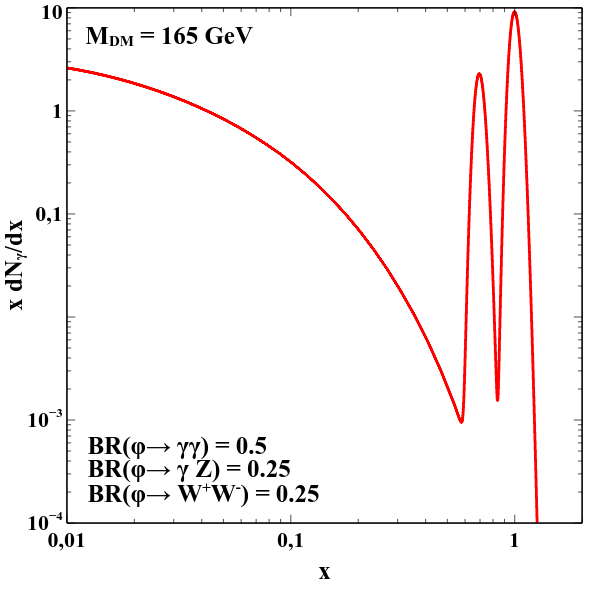} 
\caption{\label{fig:spectrum} This spectrum can only be explained with CP violation. In this case,  $M_{DM}=165~$GeV, $BR(\varphi\to \gamma\gamma)=0.5$,  $BR(\varphi\to \gamma Z)=0.25$, and $BR(\varphi\to W^+W^-)=0.25$. A $10\%$ detector energy resolution was assumed to generate this figure. Notice that the two lines are clearly distinguishable. }
\end{figure}

From the left panel, for instance, one can see that $BR(\varphi\to \gamma Z)$ has a maximun of about $60\%$ for a CP-even scalar and of about $90\%$ for a CP-odd or CP violating scalar. Thus, if the data were to show that  $BR(\varphi\to \gamma Z)$ is larger than $60\%$, we would conclude that the dark matter is \emph{not} a CP-even particle. Analogously, if $BR(\varphi\to W^+W^-)+BR(\varphi\to ZZ)$ were found to be larger than about $15\%$, we would conclude, from the right panel, that the dark matter is \emph{not} a CP-odd particle. Notice that there are several configurations where the violation of CP could be claimed (the green points), but they all require  two different branchings to be measured. For example, branching ratios of $30\%$ and $40\%$ respectively for  $BR(\varphi\to \gamma \gamma)$ and  $BR(\varphi\to \gamma Z)$ would indicate CP violation, according to the left panel.  Similarly,  if  $BR(\varphi\to \gamma \gamma)$ and  $BR(\varphi\to W^+W^-)+BR(\varphi\to ZZ)$ were  found to be respectively around $50\%$  and $25\%$, we would conclude that the dark matter interactions violate CP. Such branchings, in fact, can only  be reached with the CP violating points --see right panel. 

\begin{figure*}[tbh!]
  \includegraphics[width=0.9\columnwidth]{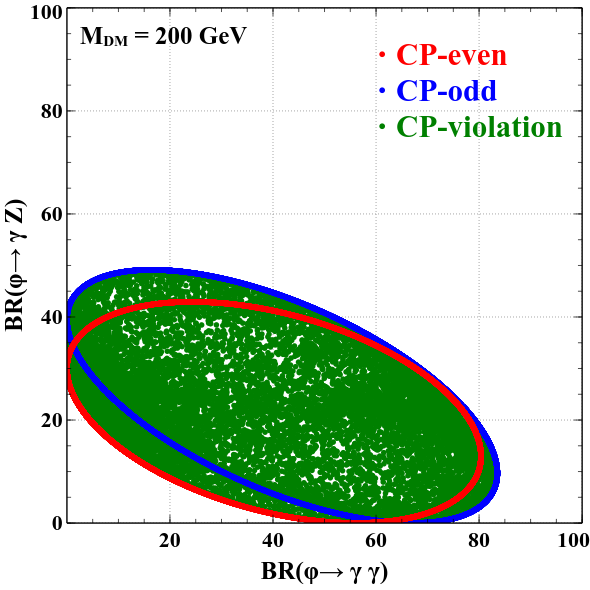}\hspace{5mm}
  \includegraphics[width=0.9\columnwidth]{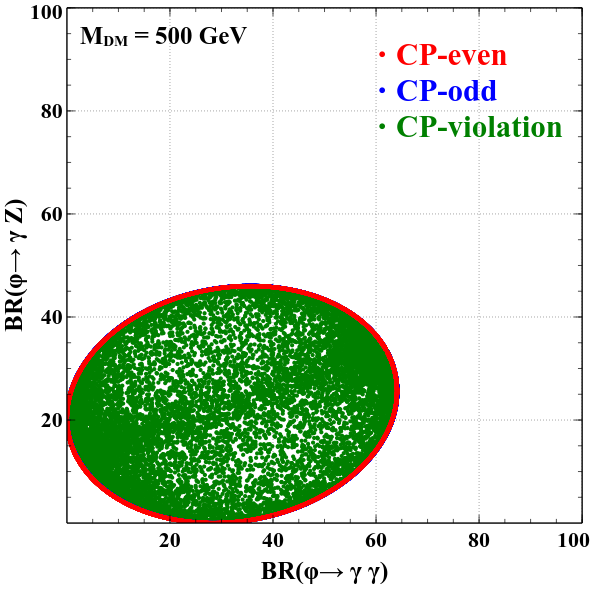} 
\caption{\label{fig:corr2} Correlations between the branching ratios into $\gamma\gamma$ and $\gamma Z$ for  $M_{DM}=200~$GeV (left panel) and $M_{DM}=500~$GeV (right panel) .}
\end{figure*}

Figure \ref{fig:spectrum} shows an example of a $\gamma$-ray spectrum that, within our setup, can only be explained with CP violation.  It corresponds to $M_{DM}=165~$GeV, $BR(\varphi\to \gamma\gamma)=0.5$,  $BR(\varphi\to \gamma Z)=0.25$, and $BR(\varphi\to W^+W^-)=0.25$.  In the figure one can clearly differentiate the two lines, from the $\gamma\gamma$ and $\gamma Z$ final states, and the continuum contribution, from $\gamma Z$ and $W^+W^-$ (we used the PPPC program \cite{Cirelli:2010xx} to get this contribution).  For this figure, we  adopted a $10\%$ energy resolution, which is the  Fermi-LAT value at $100$~GeV.  If the data were consistent with such a spectrum arising from dark matter decay, we would conclude that CP is violated by the dark matter interactions. 
%


These correlations between the different final states turn out to  strongly depend on the dark matter mass. For $\mdm<2M_W$, only the decay modes $\gamma\gamma$ and $\gamma Z$ are kinematically allowed and no variation is observed with the CP properties of the dark matter particle.  Both branchings simply vary between $0$ and $1$ satisfying  $BR(\varphi\to \gamma \gamma) + BR(\varphi\to \gamma Z)=1$. Hence, it is only for  $\mdm>2M_W$ that the branchings can in principle tell us something about the CP nature of the dark matter. 
The left panel of figure \ref{fig:corr2} shows the correlation between $BR(\varphi\to \gamma \gamma)$ and $BR(\varphi\to \gamma Z)$  for a higher dark matter mass, $\mdm=200~$GeV. From the figure we  see that the two ellipses corresponding to the CP-even and the CP-odd points  are now closer to each other, making more difficult to differentiate between these two possibilities\footnote{In any case, the correlation with $ZZ$ and $W^+W^-$ provides additional information}. If, for instance,  branching ratios of $10\%$ and $15\%$ were measured respectively for $\gamma\gamma$ and $\gamma Z$,  a CP-odd particle would be excluded.  Confirming CP violation, on the other hand, continues to be equally feasible for this higher mass.  One example of a CP-violating point is  $BR(\varphi\to \gamma \gamma)=40\%$ and $BR(\varphi\to \gamma Z)=25\%$. 

%

At even higher masses it becomes impossible, even in principle, to distinguish between the CP-even and the CP-odd cases (see appendix), but it remains feasible to  demonstrate CP violation. This situation is illustrated in the right panel of figure \ref{fig:corr2}, which shows the correlation between the $\gamma\gamma$ and $\gamma Z$ branching ratios  for $\mdm=500~$GeV. In this case, the CP-even and the CP-odd ellipses lie exactly on top of each other, meaning that the branchings have completely lost their discriminating power between these two possibilities. We still find, nonetheless, plenty of CP violating points (green) lying inside the ellipses, so CP violation could be established. The limiting factor at high masses is the capability of differentiating the two lines (a prerequisite to measure their branchings), which depends on the energy resolution of the detector.  The GAMMA-400 $\gamma$-ray telescope \cite{Topchiev:2017xfp}, which aims at $1\%$ energy resolution, would allow to distinguish the two lines up to dark matter  masses of $400~$GeV  for annihilating dark matter \cite{Bringmann:2012ez}, and slightly higher values for the case of decaying dark matter relevant to our proposal. A detailed analysis of this issue as well as of the precision in the measurement of the branchings that can be achieved in future experiments is left for future work. Another direction that is definitely worth exploring is whether it is possible to determine the CP nature of annihilating rather than decaying dark matter.


\section{Conclusion}
We have shown  that, under certain conditions, it is  possible to determine whether the dark matter particle is a scalar (CP-even), a pseudoscalar (CP-odd), or if its interactions violate CP. The basic requirement to discriminate among these possibilities is the observation of $\gamma$-ray lines arising from the \emph{decay} of a spin-zero dark matter particle.  We argued that such a decay would most naturally be explained within the framework of effective field theory and demonstrated that, as a consequence of gauge invariance, there exists correlations between the  branching ratios into $\gamma\gamma$ and $\gamma Z$ that depend on the mass and the CP nature of the dark matter particle. Thus, once these $\gamma$-ray lines are detected, one can check to see if their branchings are consistent with the predictions for a scalar, a pseudoscalar or a CP-violating scalar.   In this way, it is possible to determine one of the fundamental properties of the dark matter particle: its CP nature.

\section*{Acknowledgments}
FSQ acknowledges support from MEC, UFRN and ICTP-SAIFR FAPESP grant 2016/01343-7. We are grateful to Stefano Profumo and Yann Mambrini for comments. \vspace{2cm}
\appendix*
\section{Analytic expressions}
\label{sec:appendix}
The squared matrix elements for the two-body decays we consider can be obtained from the Lagrangian, equation (\ref{eq:Lagrangian}), and are given by
\begin{widetext}
\begin{align}
\mplanck^2\,|\mathscr{M}(\varphi\to\gamma\gamma)|^2& = 4\mdm^4\left[(b\,\cw+d\,\sw)^2+(a\,\cw+c\,\sw)^2\right]\\
\mplanck^2\,|\mathscr{M}(\varphi\to\gamma Z)|^2& = 8\cw\sw(\mdm^2-M_Z^2)^2\left[(b-d)^2+(a-c)^2\right]\\
\mplanck^2\,|\mathscr{M}(\varphi\to W^+W^-)|^2& = 8\left[\mdm^4\left(c^2+d^2\right)-4\mdm^2M_W^2\left(c^2+d^2\right)+6c^2M_W^4\right]\\
\mplanck^2\,|\mathscr{M}(\varphi\to Z Z)|^2& = 4\left[\mdm^4\left((b\,\sw+d\,\cw)^2+(a\,\sw+c\,\cw)^2\right)\right.\\\nonumber
&~~-4\mdm^2M_Z^2\left((b\,\sw+d\,\cw)^2+(a\,\sw+c\,\cw)^2\right)\\\nonumber
&~~\left.+6M_Z^4(a\,\sw+c\,\cw)^2\right],
\end{align}
\end{widetext}
where $\sw=\sin\theta_W$, $\cw=\cos\theta_W$, and $\mdm$ denotes the mass of the dark matter particle.  It is straightforward to compute from these expressions the decay width into the different channels and the corresponding branching ratios. From them, indirect detection limits on the couplings $a,b,c,d$ can be derived --see e.g. \cite{Mambrini:2015sia}.

For $M_{DM}\gg M_Z$, the expressions for the branching ratios  get simplified  to
\begin{widetext}
\begin{align}
& BR(\varphi\to\gamma\gamma)=\frac{\left(a^2+ b^2\right) \cwf + 2 (a c+bd)\, \sw \cw +\left(c^2+ d^2\right) \swf} {a^2+ b^2+3\left(c^2+ d^2\right)}  \\
& BR(\varphi\to\gamma Z)=\frac{\sin^2(2\theta_W) \left((a-c)^2+ (b-d)^2\right)}{2\left(a^2+ b^2+3 \left(c^2+ d^2\right)\right)}\\
& BR(\varphi\to W^+W^-)=\frac{2 \left(c^2+ d^2\right)}{a^2+ b^2+3 \left(c^2+ d^2\right)}\\
& BR(\varphi\to ZZ)=\frac{4 \cos (2\theta_W) \left(-a^2- b^2+c^2+ d^2\right)+\cos (4 \theta_W) \left((a-c)^2+ (b-d)^2\right)}{8 \left(a^2+ b^2+3 \left(c^2+d^2\right)\right)}\\ \nonumber
  &\hspace{2.5cm}+\frac{3a^2+2 a c+3 b^2+2 b d+3 c^2+3 d^2}{8 \left(a^2+ b^2+3 \left(c^2+d^2\right)\right)}.
\end{align}  
\end{widetext}


From these results it can be checked  that the parametric dependence of the branchings is exactly the same for  the CP-even ($b=d=0$) and CP-odd ($a=c=0$) cases. Therefore, for $\mdm\gg M_Z$, the correlations between the different branching ratios cannot be used to distinguish between a scalar and a pseudoscalar dark matter particle.  This analytic result is in agreement with the numerical results illustrated in the right panel of figure \ref{fig:corr2}. Let us emphasize that it is still possible to confirm CP violation in this high mass limit.

\bibliographystyle{unsrt}
\bibliography{sample}

\end{document}